\documentclass[longbibliography,aps,prl,reprint,groupedaddress,superscriptaddress]{revtex4-1}

\usepackage[pdftex]{graphicx}
\usepackage{dcolumn}
\usepackage{bm}
\usepackage{amsmath,amssymb}
\usepackage{latexsym}

\usepackage{physics}
\usepackage{verbatim}
\usepackage{lipsum}

\usepackage{xcolor}
\definecolor{mycolor}{rgb}{0.16, 0.17, 0.57}
\usepackage{tikz}

\usepackage[T1]{fontenc}
\usepackage{hyperref}
\hypersetup{colorlinks=true, linkcolor=mycolor, citecolor=mycolor, urlcolor=mycolor,anchorcolor=mycolor}

\usepackage{pdfpages}
\makeatletter
\AtBeginDocument{\let\LS@rot\@undefined}
\makeatother

\makeindex
\begin{document}

\title{Quantum-Enhanced Sensing Enabled by Scrambling-Induced Genuine Multipartite Entanglement}
    \author{Guantian Hu}
    \thanks{These authors contributed equally to this work.}
     \affiliation{National Laboratory of Solid State Microstructures, School of Physics, Nanjing University, Nanjing 210093, China}
    \affiliation{International Quantum Academy, Futian District, Shenzhen, Guangdong 518048, China}
   
   \author{Wenxuan Zhang}
    \thanks{These authors contributed equally to this work.}
    \affiliation{International Quantum Academy, Futian District, Shenzhen, Guangdong 518048, China}
     \affiliation{Southern University of Science and Technology, Shenzhen, Guangdong 518055, China}

     \author{Zhihua Chen}
    \thanks{These authors contributed equally to this work.}
     \affiliation{School of Science, Jimei University, Xiamen, 361021,China}
    
    \author{Liuzhu Zhong}
    \affiliation{International Quantum Academy, Futian District, Shenzhen, Guangdong 518048, China}
     \affiliation{Southern University of Science and Technology, Shenzhen, Guangdong 518055, China}

   \author{Jingchao Zhao}
   \affiliation{International Quantum Academy, Futian District, Shenzhen, Guangdong 518048, China}
     \affiliation{Southern University of Science and Technology, Shenzhen, Guangdong 518055, China}

    \author{Chilong Liu}
     \affiliation{International Quantum Academy, Futian District, Shenzhen, Guangdong 518048, China}
     \affiliation{Southern University of Science and Technology, Shenzhen, Guangdong 518055, China}
       
     \author{Zixing Liu}
    \affiliation{International Quantum Academy, Futian District, Shenzhen, Guangdong 518048, China}

 \author{Yue Xu}
  \affiliation{International Quantum Academy, Futian District, Shenzhen, Guangdong 518048, China}
     \affiliation{Southern University of Science and Technology, Shenzhen, Guangdong 518055, China}

    \author{Yongchang Lin}
     \affiliation{International Quantum Academy, Futian District, Shenzhen, Guangdong 518048, China}
     \affiliation{Southern University of Science and Technology, Shenzhen, Guangdong 518055, China}

     \author{Yougui Ri}
     \affiliation{International Quantum Academy, Futian District, Shenzhen, Guangdong 518048, China}
     \affiliation{Southern University of Science and Technology, Shenzhen, Guangdong 518055, China}

     \author{Guixu Xie}
      \affiliation{International Quantum Academy, Futian District, Shenzhen, Guangdong 518048, China}
     \affiliation{Southern University of Science and Technology, Shenzhen, Guangdong 518055, China}

 \author{Mingze Liu}
  \affiliation{International Quantum Academy, Futian District, Shenzhen, Guangdong 518048, China}
     \affiliation{Southern University of Science and Technology, Shenzhen, Guangdong 518055, China}
   
 \author{Haolan Yuan}
  \affiliation{International Quantum Academy, Futian District, Shenzhen, Guangdong 518048, China}
     \affiliation{Southern University of Science and Technology, Shenzhen, Guangdong 518055, China}

 \author{Yuxuan Zhou}
     \affiliation{International Quantum Academy, Futian District, Shenzhen, Guangdong 518048, China}

\author{Yu Zhang}
 \affiliation{National Laboratory of Solid State Microstructures, School of Physics, Nanjing University, Nanjing 210093, China}

\author{Chang-Kang Hu}
\email{huchangkang@iqasz.cn}
 \affiliation{International Quantum Academy, Futian District, Shenzhen, Guangdong 518048, China}

\author{Song Liu}
\email{lius@iqasz.cn}
 \affiliation{International Quantum Academy, Futian District, Shenzhen, Guangdong 518048, China}
 \affiliation{Shenzhen Branch, Hefei National Laboratory, Shenzhen 518048, China}

\author{Dian Tan}
\email{tandian@iqasz.cn}
 \affiliation{International Quantum Academy, Futian District, Shenzhen, Guangdong 518048, China}
 \affiliation{Shenzhen Branch, Hefei National Laboratory, Shenzhen 518048, China}

\author{Dapeng Yu}
 \affiliation{International Quantum Academy, Futian District, Shenzhen, Guangdong 518048, China}
 \affiliation{Shenzhen Branch, Hefei National Laboratory, Shenzhen 518048, China}

\begin{abstract}
Quantum sensing leverages quantum resources to surpass the standard quantum limit, yet many existing protocols rely on the preparation of complex entangled states and Hamiltonian engineering, posing challenges for universality and scalability. Here, we report an experimental realization of a universal protocol, known as Butterfly Metrology, proposed in [\href{https://arxiv.org/abs/2411.12794}{arXiv:2411.12794}], demonstrating a scrambling-based approach for quantum-enhanced sensing on a superconducting quantum processor. By exploiting many-body information scrambling, we observe quantum-enhanced sensitivity to an encoded phase beyond the standard quantum limit, with a scaling consistent with a factor-of-two of the Heisenberg limit for system sizes of up to 10 qubits. 
Importantly, we experimentally establish a connection between the enhanced sensitivity and the dynamics of the out-of-time-order correlator (OTOC), and show that the buildup of scrambling-induced genuine multipartite entanglement underlies the observed sensitivity enhancement. Our results  demonstrate a scalable and practical  approach for quantum-enhanced sensing in interacting many-body quantum systems.
\end{abstract}

\maketitle

Harnessing information scrambling opens a new paradigm for quantum-enhanced sensing in many-body systems~\cite{doi:10.1126/science.adg9500,39bt-37yl}. Unlike methods that rely on static entanglement, scrambling provides a scalable approach to generate dynamic entanglement, leading to sensitivities that surpass the Standard Quantum Limit (SQL) of $1/\sqrt{N}$, where $N$ is the number of particles in the system \cite{RevModPhys.89.035002,RevModPhys.90.035006,PhysRevLett.96.010401}.
Various quantum sensing protocols often rely on the preparation of many-body entangled states, such as GHZ states \cite{doi:10.1126/science.1104149,doi:10.1126/sciadv.abg9204, doi:10.1126/science.1097576}, squeezed states \cite{PhysRevD.30.2548,PhysRevLett.116.093602,doi:10.1021/acsphotonics.9b00250, malia_distributed_2022,mao_quantum-enhanced_2023}, and NOON states \cite{doi:10.1126/science.1188172,PhysRevLett.112.103604,doi:10.1126/science.1170730}, or the ground state of many-body systems at the phase transition point~\cite{PhysRevX.8.021022,mao2023quantum,ding2022enhanced} to achieve enhanced sensitivity. 
While these approaches have proven highly successful in specific settings that rely on complex entangled-state preparation or Hamiltonian engineering, they face challenges in scalability, largely due to the stringent control requirements for preparing highly entangled states or engineering specific Hamiltonians, which become increasingly demanding in large-scale and complex many-body systems~\cite{cao2023generation,PhysRevLett.119.180511,kolodynski_efficient_2013,demkowicz-dobrzanski_elusive_2012}.

\begin{figure*}[htb]
\centering 
\includegraphics[width=1.0\linewidth]{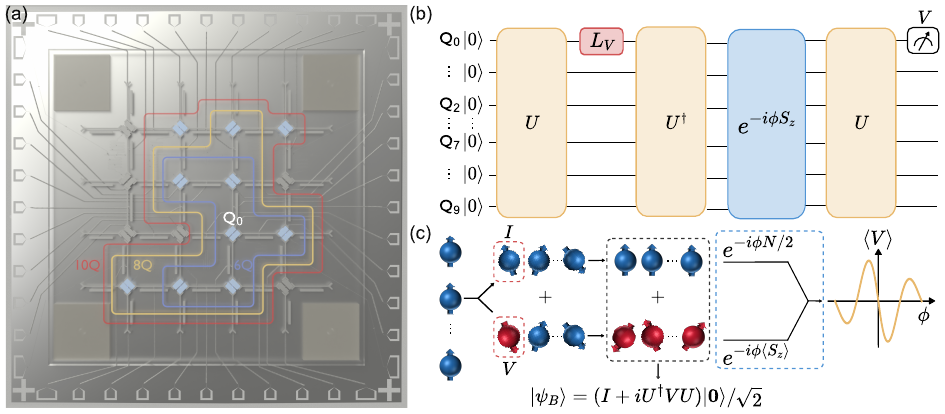}
\caption{Schematic of the butterfly metrology protocol.
(a) The superconducting quantum processor. The subsets of qubits used in the experiment (6, 8, and 10 qubits with the center qubit $Q_0$) are highlighted.
(b) Quantum circuit implementing the butterfly metrology protocol. 
(c) The protocol realizes the butterfly state, which forms a coherent superposition of two distinct branches acquiring macroscopically different phases. The resulting Ramsey-like phase accumulation of $N\phi/2$ enables quantum-enhanced sensitivity.}
\label{fig1}
\end{figure*}

Recent experimental advances in controlling and probing information scrambling in complex quantum systems, including solid-state spin \cite{doi:10.1126/science.abg5029,PhysRevLett.129.160602,PhysRevX.11.021010,braumuller2022probing,PhysRevX.7.031011}, trapped ions \cite{landsman_verified_2019,PhysRevLett.124.240505,garttner2017measuring}, or Rydberg atom arrays \cite{w1cp-l5vq}, have paved the way for scrambling-based quantum metrology. Scrambling-based protocols exploit  many-body Hamiltonian evolution with information scrambling to generate multipartite entangled states, and naturally remain effective as the system size increases. Moreover, scrambling-based protocols are universal across different Hamiltonians, allowing them to be implemented with many-body Hamiltonians and making them adaptable to a variety of quantum systems.  Together, these features make scrambling-based metrology compatible with large-scale quantum systems, where state-preparation overhead and Hamiltonian engineering pose significant challenges.
 
Notably, a fast-scrambling spin model has been experimentally implemented, demonstrating that information scrambling can be harnessed as a metrological resource~\cite{doi:10.1126/science.adg9500}. 
More recently, B. Kobrin and colleges introduced a scrambling-based quantum sensing protocol built on an entangled butterfly state \cite{kobrin2024universalprotocolquantumenhancedsensing}, establishing a universal approach for quantum-enhanced sensing. By inserting a local operation into a time-reversal sequence, the protocol exploits many-body information scrambling  to generate a highly entangled butterfly state, enabling quantum enhanced sensing with a scaling reaching a factor-of-two of the Heisenberg limit (HL).

In this Letter, we experimentally demonstrate quantum-enhanced sensing enabled by information scrambling on a superconducting quantum processor.
First of all, we explore the connection of quantum entanglement to the scrambling-enhanced sensing. We demonstrate that it is not simply the presence of entanglement, but its dynamical generation of genuine multipartite entanglement through information scrambling, that underpins the observed quantum-enhanced sensitivity. This insight clarifies how scrambling transforms local perturbations into globally accessible quantum resources, enabling enhanced sensitivity in many-body systems. Next, we  perform phase estimation and show that the sensitivity surpasses the SQL, with a scaling consistent with a factor-of-two of the HL for systems of up to 10 qubits. Furthermore, we  experimentally verify the connection between the achieved sensitivity and the OTOC, validating the scrambling-based protocol. 

\begin{figure*}[tbp]
\includegraphics[width=1\linewidth]{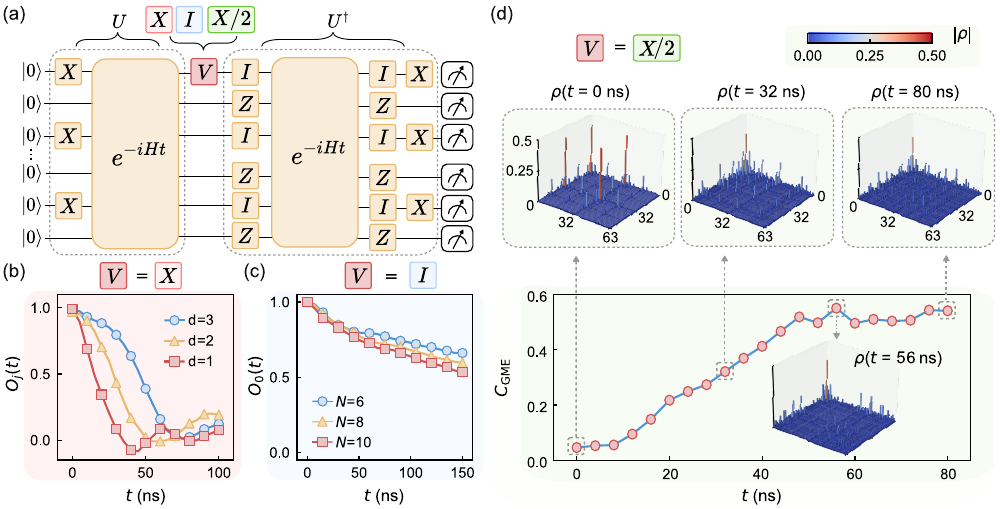} 
\caption{Characterization of the OTOC and the scrambling-induced entanglement.
(a) Quantum circuit for measuring the OTOC. 
(b) Measured $O_{j}(t)$ with $V=X$  as a function of evolution time $t$ at different qubit distance $d$ from $Q_0$. 
(c) The reference experiment measuring $O_{0}(t)$ with $V=I$, used to characterize non-ideal effects. 
(d) Dynamics of scrambling-induced entanglement quantified by the GME concurrence $C_{\mathrm{GME}}$ as a function of time $t$ for system size $N=6$ with $V=X/2$. The six-qubit density matrix is reconstructed via full quantum state tomography with a complete set of $3^6$ observables with 5000
samples for each observable measured. Insets show the reconstructed density matrices at $t=0 , 32, 56, 80$ ns.} 
\label{fig2}
\end{figure*}

\textit{Experimental setup and protocol--} Our experiment is implemented on a superconducting quantum processor that provides high controllability and scalability for quantum sensing. As shown in Fig.~\ref{fig1}(a), the processor consists of a $4\times4$ lattice of 16 superconducting qubits with nearest-neighbor connectivity.  Each qubit frequency is tunable over a range of approximately 2 $\mathrm{GHz}$. Nearest-neighbor qubits are coupled through tunable couplers that allow the precise control of the effective exchange interaction strength $J$ \cite{PhysRevLett.113.220502}. The coupling can be continuously tuned from near zero to several MHz \cite{PhysRevLett.106.060501}, allowing the system to be switched between isolated-qubit operation and interacting many-body dynamics. 
During Hamiltonian evolution, couplers are activated and all qubits are tuned into resonance, realizing a nearest-neighbor energy-exchange Hamiltonian, $H =J\sum_{\langle m,n\rangle}(\sigma_{x}^{m}\sigma_{x}^{n}+\sigma_{y}^{m}\sigma_{y}^{n})$, where 
$\langle m,n\rangle$ denotes the nearest-neighbor qubit pairs. 
In the experiment, the coupling strength is set to $J\simeq2\pi\times3$ $\mathrm{MHz}$, and the qubits are resonant at a frequency of $4.48$ $\mathrm{GHz}$. The protocol relies solely on single-qubit operations (average fidelity of $99.8\%$) combined with many-body Hamiltonian evolution, making this setup well suited for implementing scrambling-based quantum metrology. Details of the setup and protocol can be found in the supplementary materials (SM).

As illustrated in Fig.~\ref{fig1}(b), the protocol utilizes the dynamics of quantum scrambling for quantum-enhanced sensing \cite{kobrin2024universalprotocolquantumenhancedsensing}. The butterfly state $|\psi_B\rangle$ is prepared via a time-reversal sequence, in which the system evolves from a fully polarized state $|\bf{0} \rangle$ through forward evolution $U$, a local operator $L_V = (I + iV)/\sqrt{2}$ acting on the center qubit $Q_0$ (e.g. $V=X$), and subsequent backward evolution $U^\dagger$. 
As shown in Fig.~\ref{fig1}(c), the local operator $L_V$ is essential for generating the butterfly state, which creates a coherent superposition of an unscrambled branch with $\langle S_z \rangle = N/2$ and a scrambled branch $V(t)=U^\dagger VU$  with $\langle S_z \rangle \approx 0$, where $S_z =\frac{1}{2}\sum_{i=0}^{N-1} \sigma_z^{i} $
is the collective spin operator.
Since the two components carry distinct $S_z$ expectation values, encoding a Ramsey-like phase $\phi$ in $e^{-i\phi S_z}$ 
imprints in a macroscopic relative phase  $N\phi/2 $ between them,  scaling with system size $N$. 
The relative phase is subsequently retrieved by reapplying the forward evolution $U$ and measuring at observable $V$. 

\textit{Experimental results--}
The OTOC is widely used as a diagnostic of quantum information scrambling; accordingly, our protocol is directly connected to  the OTOC \cite{R2}. In our work, the measured OTOC is defined as $O_j(t)=\langle \psi | V(t) W_j V(t) W_j | \psi \rangle$, where $W_j$ is a local operator acting on qubit $Q_j$. 
In our experiment, we have $|\psi \rangle=|\mathbf{0}  \rangle $ and $W_j=\sigma_z^{j}$. Considering $\sigma_z^{j}|\mathbf{0}  \rangle=|\mathbf{0}  \rangle$ for any $j$, then the OTOC equals to $O_j(t)=\langle \mathbf{0} | V(t) \sigma_z^{j} V(t)  |\mathbf{0}  \rangle$.
As illustrated in Fig.~\ref{fig2}(a), the OTOC is measured at $Q_j$ in the $\sigma_z$ basis through the following sequence: the system starts with an initial state $|\psi\rangle=|\mathbf{0}  \rangle$, followed by the forward evolution $U$, the application of the local operator $V$, and the subsequent backward evolution $U^\dagger$. The forward evolution $U= e^{-iHt}(\prod_{k} X_k)$ was constructed as a combination of randomly applied single-qubit $X$-gates and Hamiltonian evolution $e^{-iHt}$. 
As demonstrated in Ref.\cite{braumuller2022probing}, the backward evolution $U^\dagger$ is implemented by effectively reversing the sign of the Hamiltonian $H$ through a unitary conjugation, $-H = \left( \prod_{i \in B} Z_i \right) H \left( \prod_{i \in B} Z_i \right)$,
where $B$ denotes one part of the bipartite lattice and $Z_i$ is the Pauli $Z_i$ oeprator acting on the qubit $i$. 

\begin{figure*}[tbp]
\includegraphics[width=1\linewidth]{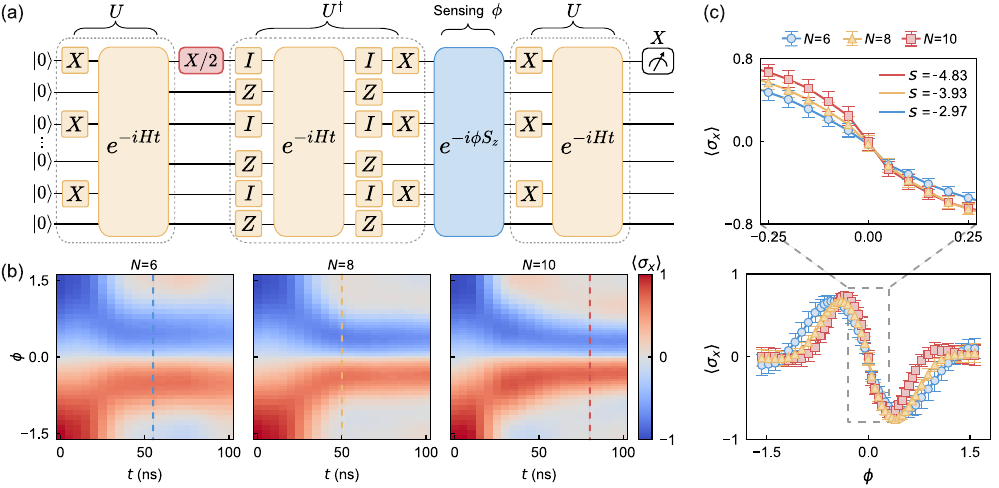} 
\caption{ Quantum sensing of the phase $\phi$. (a) The quantum circuit for sensing the phase $\phi$. (b) Measured expectation value of $\langle \sigma_x \rangle$ as a function of the phase $\phi$ and evolution time $t$ for different system sizes $N=6, 8,$ and $10$ (from left to right). Each data set is averaged over 10 experimental runs with different sets of random $X$ gates. (c) $\langle \sigma_x \rangle$ as a function of $\phi$ at the optimal evolution time with maximal slopes for different system sizes $N$. The inset shows a enlarged area of the data near $\phi=0$.} 
\label{fig3}
\end{figure*}

To probe information scrambling, we replace the local operator by $V=X$ and measure the corresponding OTOC dynamics. A clear decay of the OTOC is observed for qubit distance $d$ relative to $Q_0$ as the evolution time increases, signaling information scrambling throughout the system. Moreover, the scrambling dynamics exhibits a pronounced distance-dependent propagation of quantum information: operators acting closer to $Q_0$ result a faster decay of the OTOC, as shown in Fig.~\ref{fig2}(b).

The experiments are inevitably affected by decoherence and imperfect controls. To mitigate the resulting deviations from ideal unitary dynamics, we introduce a reference experiment to perform normalization procedure on the experimental data, following the approach used in Ref.~\cite{doi:10.1126/science.abg5029}(See SM for further details).
Here, the local operator is chosen as $V=I$. In this case, the system undergoes a forward evolution followed by its time-reversed counterpart and ideally returns to the initial state. Consequently, in the absence of information scrambling, the OTOC is expected to remain unity. As shown in Fig.~\ref{fig2}(c), the experimentally measured OTOC 
of $Q_O$ for different system size exhibits a decay, which provides a  normalization factor for the expectation values of local observable with identical evolution time.

Under Hamiltonian evolution, the butterfly state dynamically develops multipartite quantum entanglement. To characterize whether this entanglement is genuinely global, we employ the genuine multipartite entanglement (GME) concurrence $C_{\mathrm{GME}}$ \cite{ma11,chen12}, which is strictly positive for genuinely $n$-partite entangled states and vanishes for all biseparable states. For a mixed state $\rho$, the GME concurrence is defined as $C_\text{GME}(\rho)=\min_{p_i,\ket{\psi^i}}\sum_i p_i C_\text{GME}(\ket{\psi^i})$, where the minimum runs over all the pure state decompositions of $\rho=\sum_i p_i \ket{\psi^i}\bra{\psi^i}$ \cite{ma11,chen12}. Experimentally, we prepare a six-qubit butterfly state and monitor its entanglement dynamics under Hamiltonian evolution using full state quantum tomography. As shown in Fig.~\ref{fig2}(d), $C_{\mathrm{GME}}$ increases with evolution time and reaches its largest value at 56 $\mathrm{ns}$, reflecting the scrambling-induced redistribution of quantum entanglement. Crucially, it is the dynamical build-up of GME rather than its static presence that provides the entanglement resource enabling enhanced sensing, thereby highlighting the connection between information scrambling and GME.

Next, we implement the protocol of Butterfly metrology for sensing the phase $\phi$ as shown in Fig.~\ref{fig3} (a)~\cite{kobrin2024universalprotocolquantumenhancedsensing}. 
We select $Q_0$ to perform the local operator $L_V=X/2$ and measure in the $\sigma_x$ basis. 
As depicted in Fig.~\ref{fig3} (b), we measure the observable $\langle \sigma_x \rangle$ as a function of the phase $\phi$ over evolution time $t$ for different system sizes $N$ to investigate scaling behavior on sensitivity. The phase interval between the minimal and maximal values of $\langle \sigma_x \rangle$ gets narrowing until certain evolution time, indicating the increased sensitivity. 
As seen in Fig.~\ref{fig3}(c), we plot the  $\langle \sigma_x \rangle (\phi)$  for different $N$ at the maximal slope at $\phi=0$ (dash cut lines in Fig.~\ref{fig3} (b) ), defined as $s=\frac{\partial\left \langle \sigma_x \right \rangle }{\partial \phi} \mid _{\phi=0}$.
We fit the data points in Fig.~\ref{fig3} (c) using a polynomial function and extract the slope from the first derivative at $\phi=0$.
The absolute value of the slopes increase as the system size grows, e.g. $s=-2.97, -3.93, -4.83$ for $N=6,8,10$ respectively, indicating enhanced phase sensitivity with $N$. 

\begin{figure*}[tbp]
\includegraphics[width=1\linewidth]{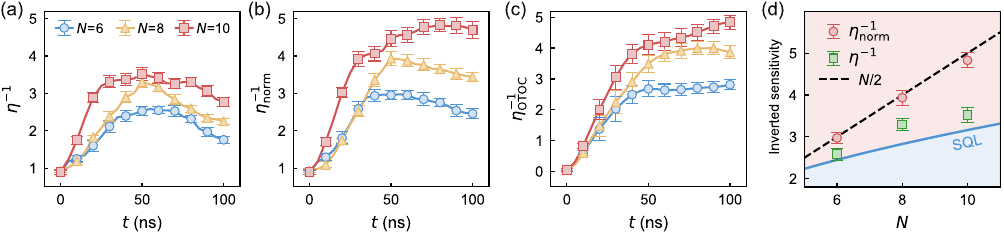} 
\caption{The sensitivity of the protocol. (a) The inverted sensitivity $\eta^{-1} $ as a function of evolution time $t$  for the system size $N=6,8,10$. (b) The inverted sensitivity $\eta_{\rm{norm}}^{-1} $ as a function of evolution time $t$  after the normalization process.  (c) The quantity  $\eta_{\rm{OTOC}}^{-1} $ as a function of evolution time $t$ is obtained from the OTOC  measurement. (d) The  maximal values of  inverted sensitivity  for different $N$. The inverted sensitivities surpass the SQL (blue solid line) for all $N$ and approach the theoretical bound of $N/2$ after the normalization porcess. The results are averaged over 10 sets of random  single-qubit $X$ gates with error bars representing the standard deviation.} 
\label{fig4}
\end{figure*}

Finally, to quantitatively characterize the quantum enhanced sensitivity, we first obtain the Fisher Information (FI), which is defined as $F=\frac{1}{1-\langle \sigma_x \rangle^2 (\phi)}(\frac{\partial \langle \sigma_x \rangle (\phi)}{\partial \phi})^2$. The FI is extracted from the measured  $\langle \sigma_x \rangle (\phi)$, reaching its maximum at $\phi=0$. The sensitivity $\eta$ of phase is then derived from $\eta = 1/\sqrt{F}$ \cite{R9,R10}. In Fig.~\ref{fig4}(a), we plot the inverted sensitivity $\eta^{-1}$ as a function of evolution time $t$  for the system size $N=6,8,10$. The $\eta^{-1}$ increases significantly as the increasing evolution time,  reaches its maximum and then decays. Taking the decoherence and imperfect controls into consideration, we apply the normalization procedure and obtain the sensitivity as displayed in Fig.~\ref{fig4}(b). The validity of this normalization is also confirmed by the numerical simulations provided in the supplementary information. 
It shows that for all different $N$,  $\eta^{-1}$ increases to a maximal value as the evolution time increases. 
This inverted sensitivity reaches its maximum approximately at $t$=56 ns for $N=6$ when the system is full scrambled with maximal GME concurrence (Fig.~\ref{fig2} (d)). 
The sensitivity is also quantitatively linked to the OTOC of the system through  $ \eta_{\rm{OTOC}}^{-1}=N/2-\sum_j O_j(t)/2$ \cite{kobrin2024universalprotocolquantumenhancedsensing}. Accordingly, we also plot the  $ \eta_{\rm{OTOC}}^{-1} $ obtained from the OTOC measurement, as shown in Fig.~\ref{fig4}(c). As the evolution time $t$ increases, information scrambling develops and the $O_j(t)$ decays from unity toward zero 
as the system gets fully scrambled. Correspondingly, the quantity $\eta_{\rm{OTOC}}^{-1}$ increases from zero and saturates near $N/2$, exhibiting the similar temporal behavior as the measured sensitivity shown in Fig.~\ref{fig4}(b). These results reveals a direct connection among quantum information scrambling, GME and quantum-enhanced sensitivity.

For classical systems, the Fisher Information limit is the SQL, $F_{\text{SQL}}=N$. 
While the quantum counterpart with enhanced sensitivity over the SQL is achieved when the experimentally measured FI exceeds $N$ \cite{PhysRevLett.104.133601}. 
As shown in Fig.~\ref{fig4}(d), the measured $\eta^{-1}$  has already exceeded the SQL,  $\eta^{-1}_{\rm{SQL}}=\sqrt{N}$. 
Moreover, after normalization, the $\eta_{\rm{norm}}^{-1}$ scales as the system size $N$, demonstrating that the Butterfly Metrology protocol can leverage scrambling dynamics to achieve sensitivity with the scaling of HL.

\textit{Conclusion.--} 
In this work, we experimentally demonstrate a scrambling-based quantum metrology protocol on a superconducting quantum processor, realizing a universal and scalable sensing approach that exploits many-body dynamics and achieves a sensitivity scaling with a factor of two of the Heisenberg limit. 
Beyond metrological performance, we experimentally reveal a direct connection among information scrambling, the emergence of genuine multipartite entanglement, and the resulting quantum-enhanced sensitivity. By combining OTOC measurements with multipartite entanglement characterization, we show that the dynamical buildup of scrambling-induced entanglement underlies the observed sensitivity enhancement. These results establish information scrambling as a quantum resource for quantum metrology with implications for practical quantum sensing in interacting many-body quantum systems.

\paragraph{Note added---} After completion of our manuscript, we note a recent work on
demonstration of Information-Scrambling-Enhanced quantum sensing beyond the standard quantum limit on a superconducting platform~\cite{ge2025informationscramblingenhancedquantumsensingstandard}.

\paragraph{Acknowledgments---}
This work is supported by the National Natural Science Foundation of China (12574550, 11934010,12004167, 12205137), the Key-Area Research and Development Program of Guangdong Province (Grants No. 2018B030326001),  the Quantum Science and Technology-National Science and Technology Major Project (Grant No. 2021ZD0301703).

\end{document}


\title{\Title}

\author{Guantian Hu}
\thanks{These authors contributed equally to this work.}
\affiliation{\nju}\affiliation{\szkl}

\author{Wenxuan Zhang}
\thanks{These authors contributed equally to this work.}
\affiliation{\szkl}\affiliation{\gdpkl}

\author{Zhihua Chen}
\thanks{These authors contributed equally to this work.}
\affiliation{\JMU}

\author{Liuzhu Zhong}
\affiliation{\szkl}\affiliation{\gdpkl}

\author{Jingchao Zhao}
\affiliation{\szkl}\affiliation{\gdpkl}

\author{Chilong Liu}
\affiliation{\szkl}\affiliation{\gdpkl}

\author{Zixing Liu}
\affiliation{\szkl}\affiliation{\gdpkl}

\author{Yue Xu}
\affiliation{\szkl}\affiliation{\gdpkl}

\author{Yongchang Lin}
\affiliation{\szkl}\affiliation{\gdpkl}

\author{Yougui Ri}
\affiliation{\szkl}\affiliation{\gdpkl}

\author{Guixu Xie}
\affiliation{\szkl}\affiliation{\gdpkl}

\author{Mingze Liu}
\affiliation{\szkl}\affiliation{\gdpkl}

\author{Haolan Yuan}
\affiliation{\szkl}\affiliation{\gdpkl}

\author{Yuxuan Zhou}
\affiliation{\szkl}

\author{Yu Zhang}
\affiliation{\nju}

\author{Chang-Kang Hu}
\email{huchangkang@iqasz.cn}
\affiliation{\szkl}

\author{Song Liu}
\email{lius@iqasz.cn}
\affiliation{\szkl}\affiliation{\HFNLSZ}

\author{Dian Tan}
\email{tandian@iqasz.cn}
\affiliation{\szkl}\affiliation{\HFNLSZ}

\author{Dapeng Yu}
\affiliation{\szkl}\affiliation{\HFNLSZ}

\maketitle
\tableofcontents
\section{Experimental setup and device parameters}
\subsection{Experimental setup}
The superconducting qubit processor is housed in a dilution refrigerator and operated at a base temperature of approximately 10 mK. The processor comprises 16 qubits arranged in a 4 $\times$ 4 lattice. Each qubit is coupled to an individual resonator for readout, with the 16 resonators distributed across two readout lines, eight per line. Nearest-neighbor qubits are connected by tunable couplers; all 24 couplers are individually controlled through dedicated flux lines, allowing flexible control of the coupling strength.

\begin{figure*}
    \centering
    \includegraphics[width=1\linewidth]{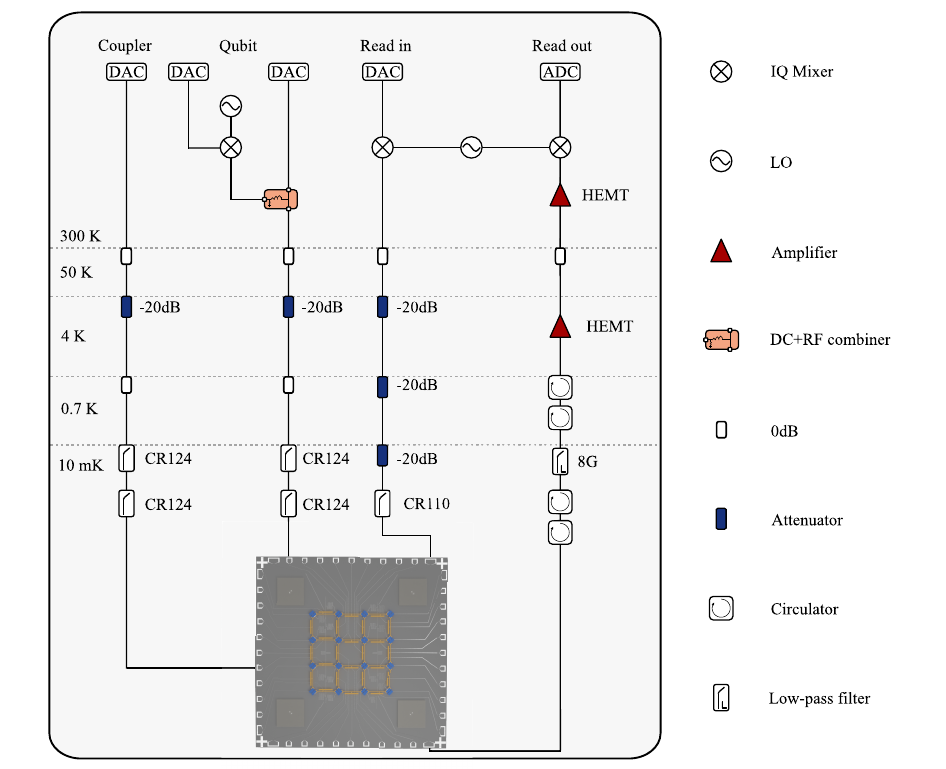}
    \caption{Schematic diagram of the setup.}
    \label{s1}
\end{figure*}

Inside the dilution refrigerator, 
the processor is equipped with   XYZ control lines for 16 qubits, 
individual flux lines for  24 couplers, two input lines and two output lines for qubit readout.
Attenuators are placed at various temperature stages for thermalization and noise attenuation. At the 10 mK stage, each control line is equipped with two CR124 filters to effectively remove unwanted noise at the qubit operating frequency. In addition, the output lines are equipped with circulators and low-pass filters at  the 10 mK stage to block reflected signals and noise  from higher temperature stages. The output signal is amplified by a high electron mobility transistor (HEMT) amplifier at the 4 K stage, followed by further amplification with a room temperature amplifier. A detailed overview of the wiring and electronics setup is presented in Fig.~\ref{s1}. 

The qubit control signals consist of high-frequency microwave signals for XY pulses and  low-frequency signals for frequency tuning, which are combined using a duplexer and delivered to the processor via the XYZ control lines. 
The couplers are controlled directly using low-frequency signals through the flux lines. 
High-frequency signals for qubit control and readout are respectively generated by up-converting intermediate-frequency (IF) signals from an AWG using a local oscillator (LO). Low-frequency signals for tuning the qubit and coupler frequencies are generated directly by the AWG. The output signal is demodulated and subsequently digitized by an analog-to-digital converter (ADC) in  Zurich Instrument UHFQA .

\subsection{Device parameters}
Table~\ref{tab1} summarizes the key parameters of the ten superconducting qubits used in the experiment, as detailed below:
\begin{itemize}

\item Qubit maximum frequency $\omega_{q}^{\rm{max}}$ (sweet-point): the values of maximum qubit frequencies are set to alternatively high and low  in order to minimize the crosstalk among the qubits.
\item Qubit idle frequency $\omega_{q}^{\rm{idle}}$: the qubit idle frequency is selected based on three considerations. First, it is kept close to the maximum frequency to reduce dephasing and maintain the dephasing time $T_2$. Second, frequencies corresponding to parasitic two-level systems (TLS) are avoided to preserve a long energy relaxation time $T_1$. Finally, the frequency detuning between nearest-neighbor and next-nearest-neighbor qubits is chosen to be much larger than the coupling strength, and the detuning between nearest-neighbor qubits is kept far from the qubit anharmonicity to suppress leakage to higher energy levels.
\item Qubit anharmonicity: $\alpha = \omega_{12}-\omega_{01}$, where $\omega_{ij}$ denotes the transition frequency from state i to state j.
\item Readout resonator frequency $\omega_r$: the readout resonators are designed with distinct frequencies for each qubit, with frequency spacings on the order of several tens of megahertz.
\item Readout fidelity: the readout fidelities $f_{gg}$ and $f_{ee}$ correspond to the the $|0\rangle$ and $|1\rangle$ states, respectively. The average readout fidelities are 0.953 for $|0\rangle$ and 0.912 for $|1\rangle$.
\item $T_1^{\rm{idle}}$: energy relaxation time at idle point,  the average $T_1^{\rm{idle}}$ is 37.8 $\mu$s.
\item $T_2^{\rm{idle}}$: dephasing time at idle point,  the average $T_2^{\rm{idle}}$ is 6.5 $\mu$s.

\end{itemize}

\begin{table}[t]
  \centering
  \setlength{\tabcolsep}{6pt}
  \renewcommand{\arraystretch}{1.25}

  \caption{Parameters of the superconducting quantum processor.}
  \label{tab1}
  \begin{tabular}{l *{10}{c}}
    \toprule
    & $Q_{0}$ & $Q_{1}$ & $Q_{2}$ & $Q_{3}$ & $Q_{4}$ & $Q_{5}$ & $Q_{6}$ & $Q_{7}$ & $Q_{8}$ & $Q_{9}$ \\
    \midrule
    $\omega_q^{\rm{max}}/2\pi\;(\mathrm{GHz})$
    & 4.624 & 4.759 & 4.536 & 4.751 & 4.536 & 4.734 & 4.825 & 4.692 & 4.527 & 4.791 \\
    $\omega_q^{\rm{idle}}/2\pi\;(\mathrm{GHz})$
    & 4.480 & 4.705 & 4.443 & 4.746 & 4.473 & 4.718 & 4.819 & 4.668 & 4.428 & 4.782 \\
    $\alpha/2\pi\;(\mathrm{MHz})$
    & -190 & -184 & -192 & -167 & -183 & -193 & -190 & -180 & -186 & -179 \\
    $\omega_r/2\pi\;(\mathrm{GHz})$
    & 6.267 & 6.425 & 6.307 & 6.484 & 6.339 & 6.543 & 6.455 & 6.496 & 6.316 & 6.513 \\
    $f_{gg}$
    & 0.959 & 0.954 & 0.943 & 0.949 & 0.938 & 0.959 & 0.964 & 0.959 & 0.953 & 0.954 \\
    $f_{ee}$
    & 0.939 & 0.931 & 0.904 & 0.914 & 0.915 & 0.910 & 0.887 & 0.917 & 0.913 & 0.886 \\
    $T_{1}^{\mathrm{idle}}\;(\mu\mathrm{s})$
    & 33.9 & 24.5 & 47.9 & 37.7 & 31.1 & 45.2 & 29.7 & 39.4 & 57.3 & 31.1 \\
    $T_{2}^{\mathrm{idle}}\;(\mu\mathrm{s})$
    & 12.2 & 8.8 & 4.5 & 6.1 & 4.6 & 4.8 & 4.0 & 11.0 & 3.7 & 5.8 \\
    \bottomrule
  \end{tabular}
\end{table}

\section{Experimental calibration}
\subsection{Flux pulse distortion calibration}

Precise calibration of distortions in the flux pulse is essential to achieve high-fidelity and stable quantum operations. Distortion in the applied flux pulse typically arises from impedance mismatches, leading to unintended qubit-frequency excursions and the accumulation of phase errors. 
In this work, we adopt an improved calibration method that corrects the flux pulse distortion while being intrinsically insensitive to qubit dephasing~\cite{PhysRevLett.129.010502}.

\begin{figure*}[htb]
    \centering
    \includegraphics[width=1\linewidth]{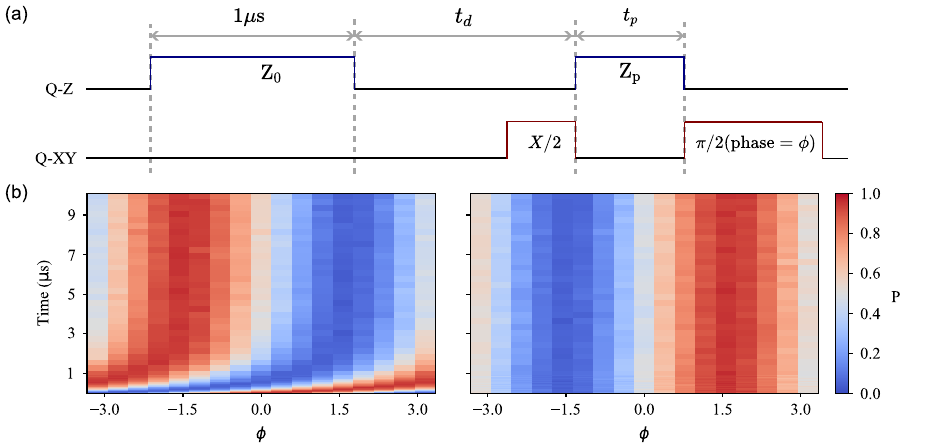}
    \caption{Flux pulse distortion calibration experiment. (a) The pulse sequence used in the experiment. $\rm{Z}_0$ and $\rm{Z}_p$ are the flux reference pulse and the flux probe pulse, respectively; $t_d$ denotes the time interval from the end of the falling edge of $\rm{Z}_0$ to the start of $\rm{Z}_p$. (b) Experimental results. The left panel shows the result without distortion calibration, and the right panel shows the result after distortion calibration. The vertical axis represents the value of $t_d$, and the horizontal axis represents the phase of the $\pi/2$ pulse.
}
    \label{s2}
\end{figure*}

The pulse sequence used in the experiment is shown schematically in Fig.~\ref{s2}(a). The qubit is maintained at its idle point when no flux pulse is applied. We first apply a 1 $\mathrm{\mu}$s-length flux pulse $\rm{Z}_0$ with a large amplitude $z_0$. Distortion of the flux pulse causes a prolonged effect on the qubit beyond the falling edge, resulting in accumulated phase errors. We then implement a Ramsey-like probe: a $X/2$ pulse is followed by a flux probe pulse $\rm{Z}_p$ with amplitude $z_{p}$ and length $t_p$, which biases the qubit to a flux-sensitive operating point to amplify its response to the residual distortion. Finally a phase-shifted $\pi/2$ pulse is applied. This protocol enables the measurement of the phase variation induced by the distorted flux pulse. The accumulated phase in this procedure can be expressed as

\begin{equation}
\varphi(t_d) = - \int_{t_d}^{t_d + t_p} \omega_{q}\!\bigl(z(t)\bigr)\,\mathrm{d}t,
\end{equation}
where $\omega_q(z)$ denotes the qubit frequency as a function of the amplitude of the flux pulse $z$; the integration interval is between the $X/2$ pulse and the phase-shifted $\pi/2$ pulse. Since the distortion is typically small, we expand the qubit frequency to first order in $z$ as

\begin{equation}
\omega_{q}(z(t))
= \omega_{q}(z_p) + D(z_p)\, z_{\rm{dist}}(t).
\end{equation}
$D(z_p)=\left.\frac{d\omega_{q}}{dz}\right|_{z=z_p}$ is the first order derivative of the qubit frequency with respect to the flux pulse amplitude evaluated at $z=z_p$, and ${z_{\rm{dist}}}(t)$ describes the distortion of the applied flux pulse. We also measured the qubit phase in the absence of $\rm{Z}_0$ as a reference. The distortion can be represented by a polynomial exponential function, ${z_{\rm{dist}}}(t)=z_0 \cdot \sum_{i} a_i e^{-t/\tau_i}$. Therefore, the qubit phase variation induced by the distortion can be expressed as 

\begin{align}
\delta\varphi(t_d)
&= - \int_{t_d}^{t_d + t_p} D(z_p)\, z_{\rm{dist}}(t)\,\mathrm{d}t \notag\\
&= z_0 D(z_p) \sum_{i} \tau_i a_i
\left( e^{-(t_d+t_p)/\tau_i} - e^{-t_d/\tau_i} \right).
\end{align}

The experimental results are shown in Fig.~\ref{s2}(b). The left panel shows the result without distortion calibration, and the right panel shows the result with distortion calibration. By fitting the data, we extract a set amplitude $a_i =$ \{-0.85\%, -1.99\%, -1.46\%, -3.56\%\} and corresponding time constants $\tau_i =$ \{1400, 460, 65.5, 14.4 \} $\mathrm{ns}$. Based on the fitted parameters, we calibrate and compensate the distortion of the input flux pulse.

\subsection{Calibration of Z gate}

\begin{figure}
    \centering
    \includegraphics[width=1\linewidth]{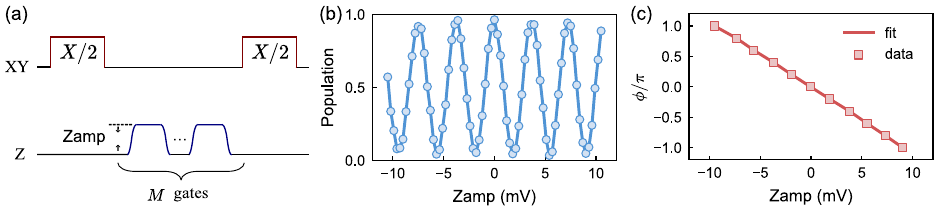}
    \caption{
Calibration of arbitrary-angle $Z$ rotation.
(a) Pulse sequence consisting of two $X/2$ gates separated by $M=5$ identical $\rm{Z}$ segments.
(b) Measured excited-state population as a function of $Z_{\mathrm{amp}}$, yielding a Ramsey fringe.
(c) The data are fitted and interpolated to determine the pulse amplitude required to implement an arbitrary target $\rm{Z}$-rotation angle.
}
    \label{s4}
\end{figure}

Accurate calibration of single-qubit $Z$ rotation is essential for implementing the backward evolution and sensing phase accumulation in our experiment. Arbitrary-angle $Z$ rotations are calibrated using a Ramsey-like sequence, as shown in Fig.~\ref{s4}(a). The sequence comprises two $X/2$ pulses with $M = 5$ identity-equivalent Z gates. By sweeping the pulse amplitude $\rm{Z}_{\mathrm{amp}}$ of the composite-$Z$ segment, we measure the excited-state population and obtain a Ramsey fringe as a function of $\rm{Z}_{\mathrm{amp}}$, as shown in Fig.~\ref{s4}(b). The measured Ramsey fringes are fitted to extract the amplitudes $\rm{Z}_{\mathrm{amp}}(\phi)$ corresponding to a discrete set of target rotation angles $\phi \in \{\pi,\,4\pi/5,\,\ldots,\,-\pi\}$. These calibration points define a mapping between the applied $Z$-pulse amplitude and the accumulated rotation angle.
To obtain a continuous calibration curve, we perform a cubic-spline interpolation of the extracted data, yielding a smooth phase-amplitude relation $\phi = f(\rm{Z}_{\mathrm{amp}})$,
as shown in Fig.~\ref{s4}(c). For any desired target rotation angle $\phi_{\rm{t}}$, this relation is numerically inverted to determine the required pulse amplitude $
\rm{Z}_{\mathrm{amp}} = f^{-1}(\phi_t)$,
which is then used to implement the corresponding arbitrary-angle $Z$ rotation in our experiment.

\subsection{Calibration of effective coupling strength}

\begin{figure}
    \centering
    \includegraphics[width=1\linewidth]{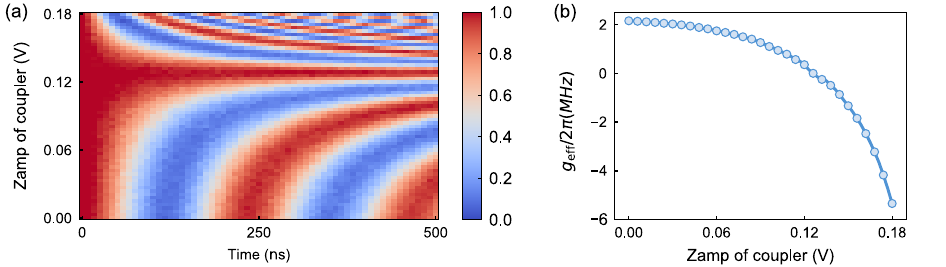}
    \caption{ Calibration of effective coupling strength.
(a) Population of the designated qubit versus $\rm{Z_{amp}}$of the coupler and evolution time.
(b) Effective coupling strength extracted from (a) as a function of $\rm{Z_{amp}}$ of the coupler.} 
    \label{iswap}
\end{figure}

We adopt the $\mathrm{Qubit\text{-}Coupler\text{-}Qubit(QCQ)}$ architecture to implement tunable coupling  between nearest-neighbor qubits.
In this architecture, the effective coupling strength is given by~\cite{YanPRA},
\begin{equation}
    g_{\rm{eff}}=g_{12}+\frac{g_{1c}g_{2c}}{2}(\frac{1}{\omega_1-\omega_c}+\frac{1}{\omega_2-\omega_c}),
\end{equation}
where $g_{12}$ denotes the direct coupling strength between two qubits, $g_{1c}$ and $g_{2c}$ represent the capacitive coupling strength between each qubit and coupler. The frequencies of the two qubits and the coupler are denoted by $\omega_1,\omega_2, \omega_c$, respectively. By tuning the coupler frequency $\omega_c$, the effective coupling strength between these qubits can therefore be continuously adjusted. As shown in Fig.~\ref{iswap}(a), one of the two qubits is initialized in the $\left | 1  \right \rangle $ state, and its excited-state population is measured while the two qubits are biased into resonance. The oscillation frequency of the population dynamics is modulated by the amplitude of flux pulse $\rm{Z_{amp}}$ applied on the coupler. By fitting the measured oscillation frequency, we extract the effective coupling strength $g_{\rm{eff}}$ as a function $\rm{Z_{amp}}$ of the coupler, as shown in Fig.~\ref{iswap}(b). The coupler frequency can then be tuned to realize a desired target coupling strength.

\section{Experimental sequence}

\subsection{Realization of time-reversed evolution}
The key challenge in our experiment is the realization of time-reversal  Hamiltonian evolution.
As illustrated in Fig.~\ref{s5}(a), 
this is achieved by conjugating the forward evolution $e^{-iHt}$ with a product of single-qubit $Z$ gates applied to a checkerboard (bipartite) subset of qubits \cite{braumuller2022probing},
\begin{equation} 
\Sigma_Z ={\textstyle \prod_{i\in \mathrm{red}}^{}} Z_i,
\end{equation}
where qubits labeled in red receive a $Z$ gate and qubits labeled in blue receive the identity gate.

The system Hamiltonian is characterized by nearest-neighbor XY coupling,
\begin{equation}
H = J \sum_{\langle m,n\rangle}\left(\sigma_x^{m}\sigma_x^{n}+\sigma_y^{m}\sigma_y^{n}\right),
\end{equation}
where $J$ is the effective coupling strength and $\langle m,n\rangle$ denotes nearest-neighbor pairs.
Using $Z\sigma_x Z=-\sigma_x$ and $Z\sigma_y Z=-\sigma_y$, one finds that for any adjacent pair
$\langle m,n\rangle$,
\begin{equation}
\Sigma_Z\left(\sigma_x^{m}\sigma_x^{n}+\sigma_y^{m}\sigma_y^{n}\right)\Sigma_Z
= -\left(\sigma_x^{m}\sigma_x^{n}+\sigma_y^{m}\sigma_y^{n}\right),
\end{equation}
Therefore, $\Sigma_Z H \Sigma_Z = -H$, and consequently,
\begin{equation}
\Sigma_Z\, e^{-iHt}\, \Sigma_Z \;=\; e^{-i(\Sigma_Z H \Sigma_Z)t} \;=\; e^{iHt}.
\end{equation}

\subsection{Pulse sequence for quantum sensing}
\begin{figure*}
    \centering
    \includegraphics[width=1\linewidth]{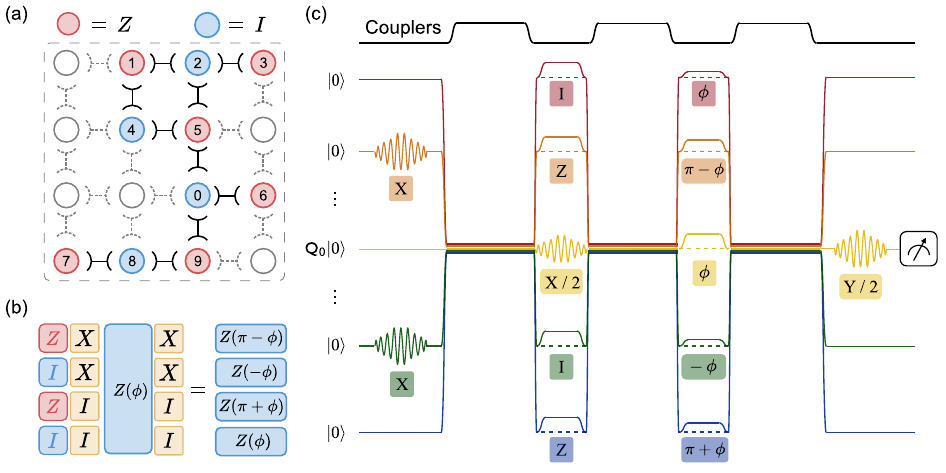}
    \caption{
Pulse sequence used in our quantum sensing experiment.
(a) Implementation of $\Sigma_Z$ by applying $Z$ gates to the red-colored qubits and $I$ gates to the blue-colored qubits, arranged in a bipartite pattern.
(b) Four distinct cases of single-qubit $Z$ rotations applied to different qubits.
(c) Detailed pulse sequence used in the experiment. The center qubit is kept at a fixed  frequency throughout the experiment, while all other qubits are tuned above or below resonance outside the Hamiltonian-evolution intervals.
}\label{s5}
\end{figure*}
The complete pulse sequence implemented in our experiment is shown in Fig.~\ref{s5}(c).
During single-qubit control operations and qubit readout, both  qubits and couplers  are biased to their idle points to suppress crosstalk between different qubits.

The system starts from the grond state $\left | \mathbf{0}   \right \rangle $, followed by  applying a set of random initial $X$ gates, denoted by $ {\textstyle \prod_{k}X_k} $. Then all qubits are simultaneously biased to resonance at $\omega/2\pi = 4.48~\mathrm{GHz}$, and the couplers are tuned closer to the qubit frequencies to activate the Hamiltonian evolution. The qubits and couplers are then returned to their idle points and a set of single-qubit $Z$ gates,
$
\Sigma_Z = \prod_{i \in \mathrm{red}} Z_i
$
is applied to the red-colored qubits. To prepare the butterfly state, an $X/2$ gate is also applied to the center qubit  $Q_0$ at this stage. Subsequently, the system undergoes a second Hamiltonian evolution for the same time.
After the second Hamiltonian evoloution, different qubits acquire distinct $Z$-axis rotation angles depending on their color (red or blue) and whether the qubit is initially excited. As illustrated in Fig.~\ref{s5}(b), four distinct single-qubit rotations 
$
Z(\pi-\phi), Z(-\phi),Z(\pi+\phi) $, and $Z(\phi)$, are applied across the processor,
where $\phi$ denotes the target phase to be sensing. Finally, the system undergoes a third period of Hamiltonian evolution, after which a $Y/2$ gate is applied to the center qubit to perform a measurement in the $\sigma_x$ basis.

Throughout the experiment, in order to stabilize the reference phase in the rotating frame of the center qubit,
we lock the frequency of the center qubit at its resonant point for the entire experimental sequence. In addition,  the tunable couplers cause a weak shift in the frequency of center qubit during Hamiltonian evolution.
To compensate for this residual frequency shift, we apply an additional calibrated $Z$ pulse to the center qubit during  evolution.
As a result, the frequency of the center qubit is effectively frozen throughout the entire experiment, ensuring a fixed reference phase and avoiding phase errors in our experiment.

\section{Data processing}
\subsection{Readout calibration}

\begin{figure*}
    \centering
    \includegraphics[width=1\linewidth]{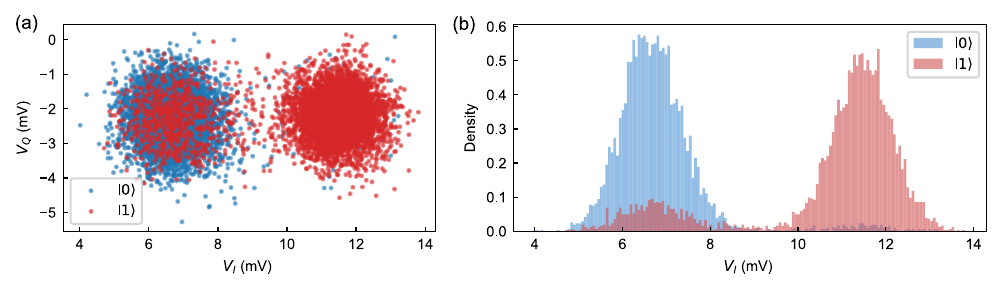}
    \caption{Single-shot measurement results. (a) IQ clusters and (b) corresponding histograms for the $|0\rangle$ and $|1\rangle$ states}
    \label{fig:iq}
\end{figure*}

In our experiment, we sweep both the frequency and power of the readout signals  to maximize the assignment fidelity for the ground and excited states. In addition, the ADC demodulation delay and demodulation window are optimized to improve the readout fidelity. Finally, readout fidelity $f_{gg}=0.959$ and $f_{ee}=0.939$ is achieved without the use of a Josephson parametric amplifier. The single-shot IQ clusters and the corresponding histograms for the $\ket{0}$ and $\ket{1}$  states are presented in Fig. \ref{fig:iq}.

Due to imperfect qubit readout, raw measurement outcomes are affected by state-dependent assignment errors. To mitigate these errors, we apply a single-qubit readout calibration based on the  readout fidelity.
The readout fidelity defines the single-qubit assignment matrix
\begin{equation}
M =
\begin{pmatrix}
f_{gg} & 1 - f_{ee} \\
1 - f_{gg} & f_{ee}
\end{pmatrix},
\end{equation}
which relates the measured probability vector $\mathbf{p}_{\mathrm{meas}}$ to the calibrated probability vector $\mathbf{p}_{\mathrm{cali}}$ via
\begin{equation}
\mathbf{p}_{\mathrm{cali}} = M^{-1} \mathbf{p}_{\mathrm{meas}}.
\end{equation}

\subsection{Normalization procedure for the experimental data}
\begin{figure*}
    \centering
    \includegraphics[width=1\linewidth]{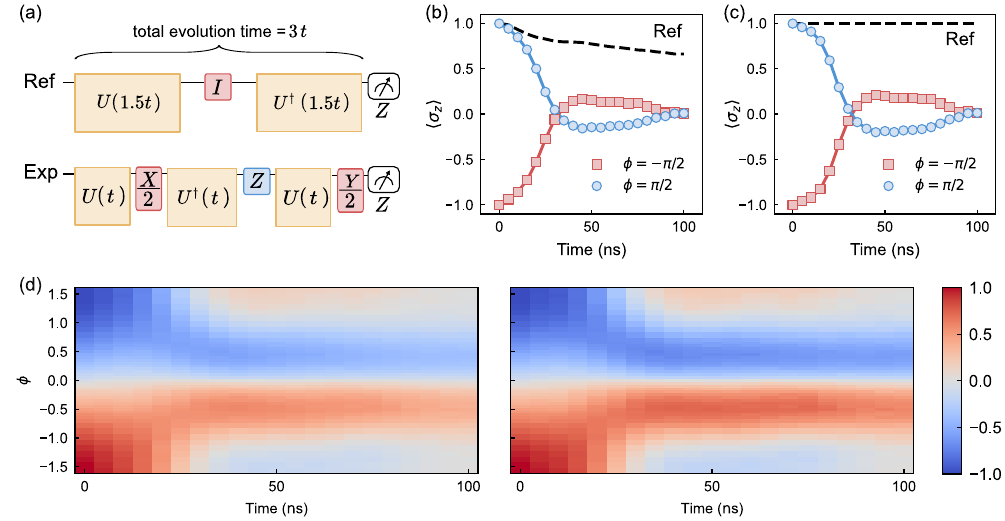}
    \caption{Experiment result without and with normalization procedure.}
    \label{mit}
\end{figure*}

The experimental results are inevitably affected by quantum decoherence and imperfect controls. To mitigate deviations arising from these non-ideal effects, we introduce a reference experiment (Ref) to perform normalization procedure on the raw experimental data (Exp)~\cite{doi:10.1126/science.1097576}, as illustrated by the circuit in Fig. \ref{mit}(a).

In the reference experiment, the gate operations applied between the two evolution segments are replaced by an identity operation $I$. Under ideal noise-free conditions, since $U^\dagger$ is the adjoint of $U$, the system return to its initial state after the sequential evolution of $U$ and $U^\dagger$. Consequently, the expectation value of the observable $\langle \sigma_z \rangle$ remains equal to unity, independent of the evolution time $t$. Any decay of $\langle \sigma_z \rangle$ observed in the reference experiment therefore directly reflects the impact of non-ideal effects on the system as shown in Fig. \ref{mit}(b).

In the primary experiment, specific gates (such as $X/2$ and $Z(\phi)$ gates shown in the figure) are applied between the two evolution stages. To compensate for the noise accumulation over the total evolution period, we normalize the experimental results using the corresponding reference data. The normalization procedure is given by:
$$\langle \sigma_{z, \text{norm}}(t) \rangle = \frac{\langle \sigma_{z, \text{exp}}(t) \rangle}{\langle \sigma_{z, \text{ref}}(1.5t) \rangle}$$
To ensure that the reference and  phase $\phi$ sensing experiment experience equivalent decoherence or imperfect controls, the evolution time of a single $U$ block in the reference circuit is set to $1.5t$, such that the total evolution duration matches that of the sensing experiment. Fig. \ref{mit}(b) presents the uncorrected raw experimental data, while Fig. \ref{mit}(c) shows the results after normalization procedure. In both cases, different values of $\phi$ are selected for demonstration.

Fig. \ref{mit}(d) displays the two-dimensional landscapes of $\langle \sigma_z \rangle$ as a function of evolution time $t$ and phase $\phi$, before and after normalization procedure. 

\begin{figure*}
    \centering
    \includegraphics[width=1\linewidth]{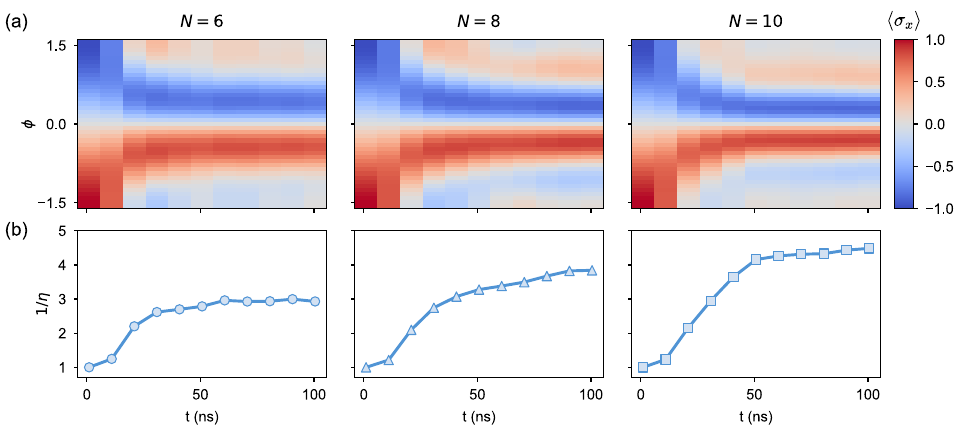}
    \caption{Numerical simulation results without noise. (a) Expectation value of the observable $\langle \sigma_x\rangle$ as a function of the sensing phase $\phi$ and the evolution time for different system sizes are obtained in simulation. (b) The corresponding inverted  sensitivity $\eta^{-1}$, extracted from the simulation data in (a), as a function of the evolution time. }
    \label{s6}
\end{figure*}

\begin{figure*}
    \centering
    \includegraphics[width=1\linewidth]{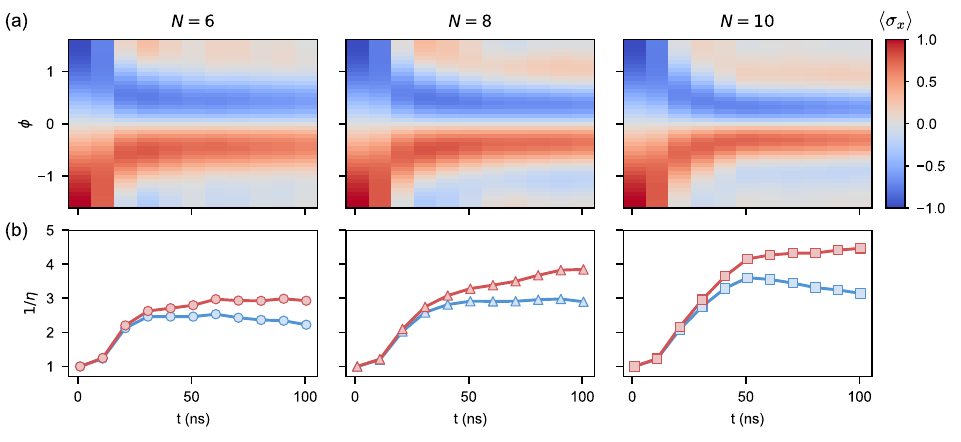}
    \caption{Numerical simulation results with noise. (a) Expectation value of the observable $\langle \sigma_x \rangle$ as a function of the phase $\phi$ and the evolution time for different system sizes $N=6,8,10$. (b) The corresponding inverted phase sensitivity $\eta^{-1}$, extracted from the simulation data in (a), as a function of the evolution time. The blue points show the raw sensitivity obtained in the presence of noise, while the red points correspond to the results after applying the normalization procedure described in Section~SIV.B. The normalization recovers the sensitivity to values consistent with the noiseless simulations as shown in Fig.~\ref{s6}(b).}
    \label{s7}
\end{figure*}

\section{Theory and simulation}
\subsection{Theory of butterfly metrology}
In this section, we theoretically derive the phase sensitivity of the butterfly metrology protocol. As illustrated in Fig. 1(c) of the main text, by combining a forward time evolution with a backward time evolution and inserting a local rotation in between, we prepare a coherent superposition of a fully polarized state and a scrambled state, which we refer to as the butterfly state $|\psi_B\rangle$. The butterfly state is given by
\begin{equation}
|\psi_B\rangle = \frac{1}{\sqrt{2}}\bigl(I + iV(t)\bigr)\,|0\rangle ,
\end{equation}

where $V(t)=U^\dagger VU$ and $V=\sigma_i^x$. A global phase encoding the parameter $\phi$ is then applied to the butterfly state, followed by another forward evolution to read out the amplified signal. The final state of the system is
\begin{equation}
|\psi\rangle = U e^{-i\phi S_z} |\psi_B\rangle .
\end{equation}

By measuring the observable $\langle V\rangle$, we obtain
\begin{equation}
\begin{aligned}
\langle V\rangle
&= \frac{1}{2}\,\left(\langle 0|V(t)|0\rangle
-\langle 0|V(t)e^{i\phi S_z}V(t)e^{-i\phi S_z}V(t)|0\rangle \right) \\
&\quad + \mathrm{Im}\!\left[
e^{i\phi N/2}\,\langle 0|V(t)e^{-i\phi S_z}V(t)|0\rangle
\right] .
\end{aligned}
\end{equation}

We approximate this evolution by a Haar-random unitary. Under this approximation, the expectation value of the first term is vanishes, while only the second term is remains nonzero. We therefore focus on analyzing the second term. For
$|\psi(t)\rangle = V(t)|0\rangle$,we expand it in the computational basis as
\begin{equation}
|\psi(t)\rangle = \sum_{s\in\{0,1\}^N} c_s |s\rangle .
\end{equation}
We define the total polarization and its probability distribution as
\begin{equation}
S_z = \sum_i \frac{s_i}{2}
\in \left\{-\frac{N}{2}, -\frac{N}{2}+1, \ldots, \frac{N}{2}\right\} .
\end{equation}

\begin{equation}
P(S_z) = \sum_{|s|=2S_z} |c_s|^2 .
\end{equation}

The expectation value $\langle V\rangle$ can then be written as
\begin{equation}
\langle V\rangle = \mathrm{Im}\!\left[
e^{i\phi N/2}\sum_{S_z} e^{-i\phi S_z} P(S_z)
\right] .
\label{eq:11}
\end{equation}

To quantify the phase sensitivity, we define
\begin{equation}
\eta_{\phi=0}^{-1}
\equiv \left.\frac{\partial \langle V\rangle}{\partial \phi}\right|_{\phi=0} .
\end{equation}

Performing a Taylor expansion of the Eq.~\eqref{eq:11} around $\phi=0$, we obtain
\begin{equation}
\langle V\rangle
\simeq \phi\left(\frac{N}{2} - \sum_{S_z} S_z P(S_z)\right) .
\end{equation}

Within the Haar-random unitary approximation, the mean of the polarization distribution is zero, yielding
\begin{equation}
\eta_{\phi=0}^{-1} = \frac{N}{2} .
\end{equation}

Alternatively, noting that
$\sigma_i^z |0\rangle = |0\rangle$,
the sensitivity can also be written in the following form:
\begin{equation}
\eta_{\phi=0}^{-1}
= \frac{1}{2}\sum_i \left[
  1 - \langle 0|V(t)\sigma_i^z V(t)\sigma_i^z |0\rangle
\right],
\end{equation}
where $\langle 0|V(t)\sigma_i^z V(t)\sigma_i^z |0\rangle$
is precisely a local OTOC. Hence, we establish a direct connection between the phase sensitivity of the butterfly protocol and information scrambling in the system.

\subsection{Numerical simulations}
We numerically simulate the measurement sensitivity in both noiseless and noisy scenarios. The results of the noiseless simulations are shown in Fig.~\ref{s6}. Figure~\ref{s6}(a) displays the observable $\langle \sigma_x\rangle$ as a function of the evolution time and the sensing phase for different system sizes ($N = 6, 8, 10$). The simulations show that the inverted sensitivity $\eta^{-1}$ increases with the evolution time. The sensitivity eventually saturates once the system reaches the full scrambling regime. In addition, the numerical results indicate that the sensitivity increases as the system size increases. As shown by the blue curves in Fig.~\ref{s6}(b), after the system reaches full scrambled, the saturation values of the inverted sensitivity $\eta^{-1}$ are consistent with the theoretical prediction of $N/2$ for all system sizes considered. The numerical simulation in the presence of noise is shown in Fig.~\ref{s7}. For short evolution times, the inverted sensitivity $\eta^{-1}$ increases as expected. However, as the evolution time increases further, the values begins to degrade due to decoherence and dephasing effects, as indicated by the blue solid curve in Fig.~\ref{s7}(b). To mitigate this degradation, we implement an normalization procedure described in Section~SIV.B,  obtaining a corrected sensitivity as shown by the red curve in Fig.~\ref{s7}(b). We find that the corrected sensitivity agrees well with the noiseless numerical results as shown in the blue curves of Fig.~\ref{s7}(b), which demonstrates the validity of the normalization procedure.